\documentclass[traditabstract,a4]{aa}  

\usepackage{epsfig}
\usepackage{graphicx}
\usepackage{epstopdf}
\usepackage{color}  
\usepackage{array}   
\usepackage{units}  
\usepackage[breaklinks, colorlinks, citecolor=blue]{hyperref}
\usepackage{natbib}  
\usepackage{multirow}   
\usepackage{amsmath,amssymb}  
\usepackage[english]{babel}
\usepackage[latin1]{inputenc}
\usepackage[T1]{fontenc}
\usepackage{longtable,lscape}
\usepackage{footnote}
\usepackage{txfonts}
\usepackage[toc,page]{appendix} 

\begin{document}

\newcommand{\avg}[1]{\left< #1 \right>}


   \title{Constraints on the dark energy dipole from large-scale structures.} 

   \author{G. Hurier
          }

\institute{Centro de Estudios de F\'isica del Cosmos de Arag\'on (CEFCA),Plaza de San Juan, 1, planta 2, E-44001, Teruel, Spain\\
\\
\email{ghurier@cefca.es} 
}

   \date{Received /Accepted}
 
   \abstract{The high-significance measurement of large-scale structure signals enables testing the isotropy of the Universe. The measurement of cosmological parameters through the large-scale distribution of matter is now a mature domain. This approach is mainly limited by our knowledge of astrophysical processes that are used to observe the large-scale structure. However, when we assume that these astrophysical processes are the same across the Universe, then it is possible to tightly constrain the isotropy of cosmological parameters across the sky. Particularly the X-SZ cross-correlation has been shown to be a probe of the large scale
structures that has
a high signal-to-noise ratio and low bias. For this analysis, we used a localized measurement of the X-SZ cross-correlation as a test of the cosmological parameter isotropy. Using the scatter of the X-SZ cross-correlation across the sky, we derive cosmological constraints $\sigma_{8} \left( \Omega_{\rm m}/0.28 \right)^{0.34} = 0.78 \pm 0.02$ and tight isotropy constraints on the dark energy dipole $\Delta \Omega_{\Lambda} < 0.07$ at 95\% confidence level.}


   \maketitle


\section{Introduction}
The cosmological principle is a pillar of current cosmological models (e.g., ${\rm \Lambda}$CDM). It assumes that the Universe is homogeneous and isotropic
at large enough scales  \citep{dod03}.\\
The cosmic microwave background (CMB) is consistent with statistical isotropy \citep{planckcosmo,planckiso}, but there are a few indications of anomalies with respect to ${\rm \Lambda}$CDM prediction \citep[see, e.g.][]{sch16}.\\

Some observations might imply that the Universe deviates from statistical isotropy \citep[see][for a review]{per14}.
Even if the Universe seems to be isotropic at the time of the recombination, it is possible to induce late-time anisotropy for example through anisotropies in the dark energy distribution. Thus, testing the isotropy of the local Universe is a powerful probe of potential dark energy anisotropies.\\
Previous analyses have tested the isotropy of the Universe using the type I supernovae (SNIa) to measure the large-scale variation of the distance modulus across the sky \citep[see, e.g.,][]{ant10,ben15,lin16}, galaxy cluster catalogs \citep[][]{ben15b,dem16}, the local velocity flow \citep{koc05}, or the X-ray background dipole \citep{pli99}.
Previous works derived $\Delta \Omega_{\rm m} /\Omega_{\rm m} < 0.43 \pm 0.06$ \citep{ant10}. Supernovae are a powerful probe of isotropy because they enable the measurement of the Universe expansion rate toward a large number of line sight.\\
The isotropy of the Universe can also be tested using the large-scale structure matter field. The number of galaxy clusters in a given volume strongly depends on the cosmological parameters.
There are two approaches to derive cosmological parameters from the large-scale structure distribution: (i) We can use the number count of galaxy clusters and directly compare it with the mass
function \citep[see e.g.,][]{tin08} calibrated on numerical simulations. (ii) We can use correlation functions. Both approaches present advantages and drawbacks: (i) The number count is highly sensitive to the selection function and the purity of the considered sample of objects, but directly allows us to explore the mass function in the mass-redshift plane. (ii) Correlation functions can be directly applied to signal maps, but they imply that we need
to consider a redshift-projected signal and contamination in the maps produced by other astrophysical objects or instrumental systematic effects.\\
Measurement of cosmological parameters through the large-scale structure distribution is now limited by our knowledge of astrophysical processes and not by statistical uncertainties. Especially the thermal Sunyaev-Zel'dovich (tSZ) effect and the X-ray emission cross-correlation has been detected at very high signal-to-noise
ratio (S/N) \citep{hur15} and is less affected than autocorrelation angular power spectra obtained with other astrophysical processes and instrumental systematics.
The tSZ effect produces a small spectral distortion in the blackbody spectrum of the CMB \citep{sun69,sun72}. Its intensity is related to the integral of the pressure along the line of sight. Its spectral signature can be used to isolate this effect from other emission on the microwave and submillimeter sky.\\
The ionized gas in the intracluster medium also produces an X-ray emission through bremsstrahlung. This radiation is proportional to the square of the electron density.\\
Considering that full-sky catalogs of galaxy clusters derived from tSZ \citep[e.g.,][]{psz2} and X-ray \citep[e.g.,][]{pif11} have complex selection functions that strongly vary across the sky, we chose to use tSZ and X-ray full-sky maps. The choice of using cross-correlation instead of autocorrelation is motivated because we aim to ensure a better control over systematic effects.\\
The paper is organized as follows: Sect.~\ref{sec_data} briefly
presents the data, then Sect.~\ref{sec_meth} describes the approach we used to perform a localized estimation of the X-SZ cross-correlation, and finally Sect.~\ref{sec_res} presents the derived constraints on cosmological parameters and isotropy.   

\section{Data}
\label{sec_data}

We made use of a MILCA tSZ fullsky map \citep{hur13a} at a resolution of 7 arcmin FWHM. This map has been obtained using Planck data \citep{planck15} from 100~to~857~GHz. In addition to the tSZ Compton parameter, the main residuals in these maps are radio point sources and high-redshift cosmic infrared background \citep[CIB,][]{planckSZCIB}.

We also used the ROSAT all-sky survey (RASS) public data\footnote{ftp://ftp.xray.mpe.mpg.de/rosat/archive/}, which cover 99.8\% of the sky, including 97\% that have an exposure time longer than 100s \citep{vog99}.
Then, we constructed a full-sky map of the photon count rate in the energy range 0.5-2.0 keV from each ROSAT photon event file and exposure map. Lower energy photons were not considered to reduce the impact of neutral hydrogen absorption that may induce some systematic anisotropy at large scales if not considered properly.
Finally, we projected each event and the associated exposure over the sky using the {\tt HEALPix} \citep{gor05} pixelization scheme at a resolution of $N_{\rm side} = 2048$.

We constructed a mask to avoid contamination by Galactic or point source emissions in Planck data.
Considering that the {\it Planck} 857 GHz channel is a good tracer of the thermal dust emission, we chose to mask all regions that present an emission above 3 $K_{\rm CMB}$\footnote{$K_{\rm CMB}$ is defined as the unit in which a blackbody spectrum at 2.725~K is flat with respect to the frequency.} at this frequency \citep[see][for unit convention and conversion]{planckBP}. We masked all sources from the {\it Planck} compact source catalog \citep{planckPCC} detected in at least one frequency with a S/N above 5.
Therefore, we masked all regions with an exposure below 100 seconds in the RASS survey. After applying these cuts, we kept about 60\% of the sky for the analysis.

\section{Methodology}
\label{sec_meth}

\begin{figure}[!th]
\begin{center}
\includegraphics[scale=0.2]{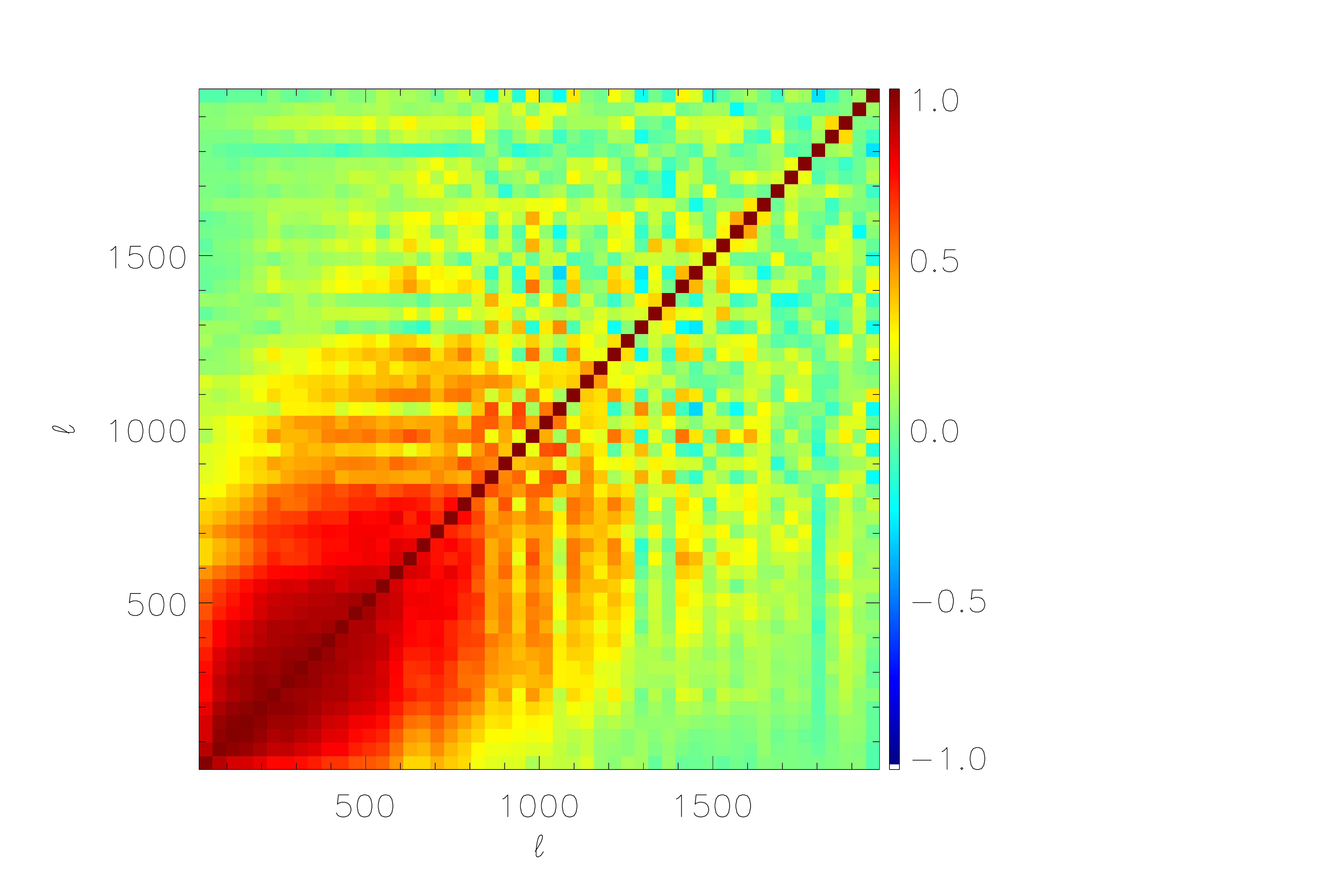}
\caption{Correlation matrix of X-SZ cross spectra computed on $\simeq 0.1$\% of the sky.}
\label{xszmat}
\end{center}
\end{figure}

The X-SZ cross power spectrum has been detected with a very high S/N ratio \citep[see][for more details]{hur15}.
Such a high-S/N probe of large-scale structures allows us to perform a localized analysis to test the isotropy of the matter field and cosmological parameter.\\
We used  {\tt HEALPix} pixelization to split the sky into 768  ($N_{\rm side} = 8$) independent sky regions that cover a sky area of roughly $7.3 \times 7.3$ degrees (corresponding to $f_{\rm sky} \simeq 0.13$\%).
Then, we rejected all regions for which the average of the mask was below 40\%, leaving $N_{\rm reg} = 416$ regions with a sufficient sky coverage.\\
We measured the X-SZ cross power spectra in each region and corrected the power spectra for mode-mixing effects induced by partial sky coverage \citep{tri05}. We computed and propagate uncertainties as presented in \citet{hur15}, and we modeled the X-SZ cross power spectrum through a halo-model approach as presented in \citet{hur14}.

From the measured cross power spectra, $C_{\ell,i}^{\rm XY}$, in a sky region $i$, we can estimate the $C_{\ell}^{\rm XY}$ covariance matrix as
\begin{align}
{\cal C}_{\ell,\ell'} = \frac{1}{N_{\rm reg}-1}\left[ \sum^{N_{\rm reg}}_i C_{\ell,i}^{\rm XY} C_{\ell',i}^{\rm XY} - \frac{1}{N_{\rm reg}} \left(\sum^{N_{\rm reg}}_i C_{\ell,i}^{\rm XY} \right) \left(\sum^{N_{\rm reg}}_i C_{\ell',i}^{\rm XY} \right)  \right].
\end{align}
Figure \ref{xszmat} presents $C_{\ell}^{\rm XY}$ correlation matrix. We observe two separate regions in this correlation matrix. At low $\ell$ ($\ell < 1000$), we observe a very high level of correlation between modes, and this level of correlation is a manifestation of the non-Gaussian X-SZ-X-SZ tri-spectrum (four-point correlation function). The amplitude of this correlation strongly depends on the number of objects that significantly contribute to the power. Consequently, as small sky fractions reduce this number of objects, it enhances the importance of these correlations.
At high $\ell$ ($\ell > 1000$), the covariance matrix is dominated by statistical uncertainty produced by the noise level in the MILCA and RASS full-sky map.

\begin{figure}[!th]
\begin{center}
\includegraphics[scale=0.15]{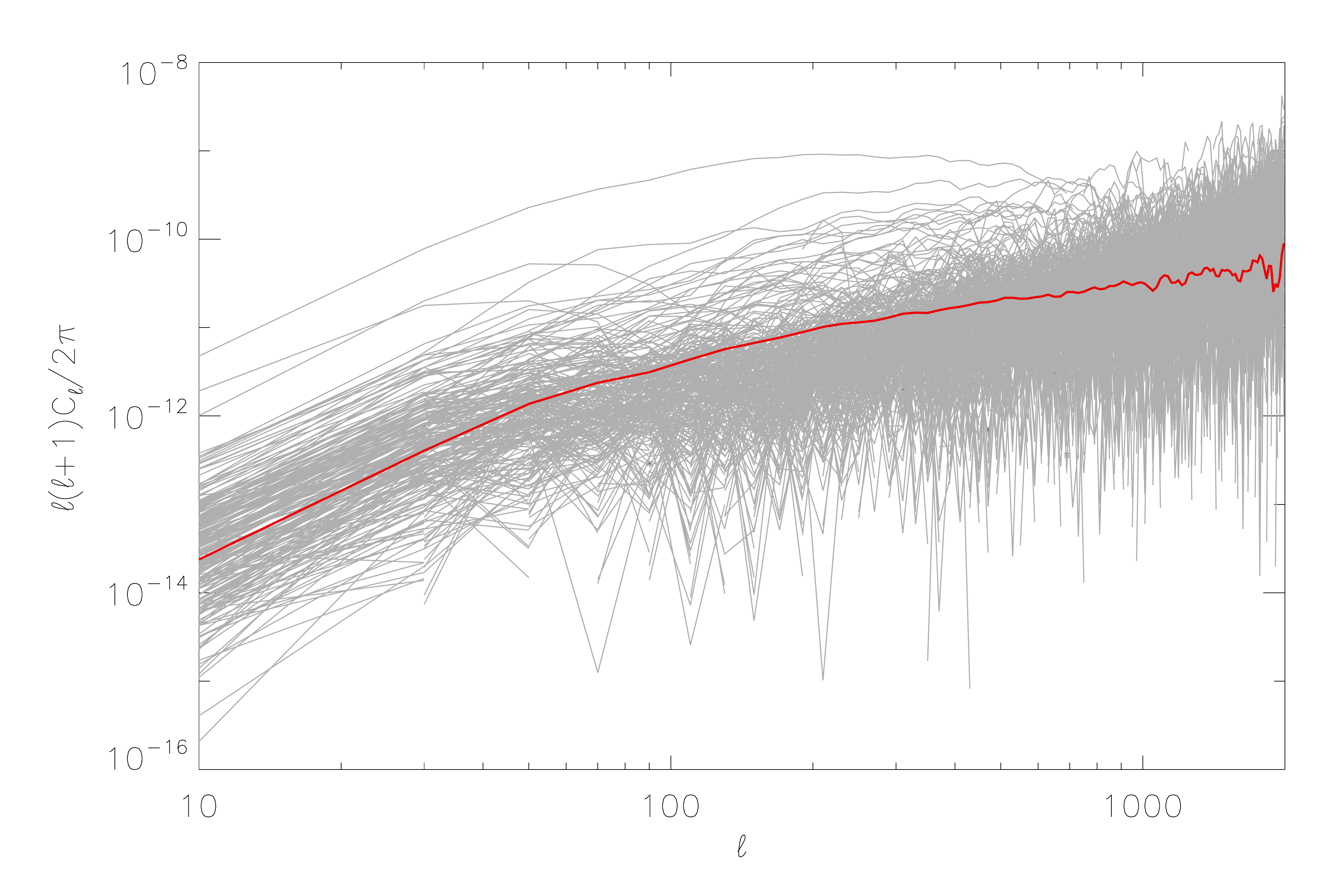}
\caption{X-SZ cross correlation power spectrum, measured on 60\% of the sky (solid red line) and measured in small sky regions, $f_{\rm sky} < 0.13$\% (solid gray lines). For readability, uncertainties are not displayed in the figure \citep[see][for the uncertainty level on the total X-SZ cross power spectrum]{hur15}.}
\label{xszsp}
\end{center}
\end{figure}

In Fig.~\ref{xszsp} we present the X-SZ cross correlation power spectra measured in independent small sky regions, as demonstrated by the study of the correlation matrix (see Fig. \ref{xszmat}). We observe a large scatter at low $\ell$ (< 1000) that is induced by the small-number statistics of bright galaxy clusters.

\section{Results}
\label{sec_res}

\subsection{Cosmological constraints from the cosmic variance}

For low-$\ell$ ($\ell < 1000$) modes, testing the isotropy of the matter field implies verifying that the observed scatter for the X-SZ cross spectrum follows a consistent statistics with what is expected from an isotropic Universe. 
As this scatter depends on cosmological parameters, testing isotropy implies fitting cosmological parameters.\\

We performed 1000 Monte Carlo simulations of the X-SZ cross power spectra for each set of cosmological parameters. 
We used Poissonian realizations of the mass function \citep{tin08} and log-normal realizations of the tSZ and X-ray fluxes consistent with the scaling laws.
We observed that the X-SZ-X-SZ tri-spectrum amplitude varies as $\Omega^{5.6}_{\rm m} \sigma^{16.5}_{8}$.\\
We note that using different order statistics does not allow us to break the degeneracies between cosmological parameters and astrophysical processes. When using different order statistics, we changed the weighting applied on the mass function. For the power spectrum, the mass function is weighted by $Y_{\rm SZ}L_{\rm X}$ , where $Y_{\rm SZ}$ and $L_{\rm X}$ are the integrated Compton parameter and the X-ray luminosity, respectively. For the X-SZ-X-SZ tri-spectrum, the mass function is weighted by $Y^2_{\rm SZ}L^2_{\rm X}$.
Thus, we are sensitive to different regions of the mass function in the mass-redshift plane that have different cosmological parameter dependencies. Consequently, the degeneracies between cosmological and astrophysical parameters are near invariant for power-spectra, bi-spectra, and tri-spectra analyses.\\
However, as power-spectra, bi-spectra, and tri-spectra analyses probe different regions of the mass function, they allow us to test the consistency of the best-fitting model for different regions of the mass-redshift plane.
In the present case, the X-SZ-X-SZ tri-spectra (measured through the scatter of the X-SZ power-spectra across the sky) leads to $\sigma_{8} \left( \Omega_{\rm m}/0.28 \right)^{0.34} = 0.78 \pm 0.02$. Similarly to previous tSZ and X-ray analyses, the uncertainties are dominated by the uncertainties over the relation between the mass of the galaxy clusters, tSZ, and X-ray fluxes. These constraints are dominated by objects at low redshift (z < 0.5) and with high mass ($M_{500} \simeq 10^{15}$ $M_{\odot}$).
This constraint is consistent with previous findings that focused on low-$z$ high-mass galaxy clusters \citep{planckSZC}.

\subsection{Constraints on the dark energy dipole}

\begin{figure}[!th]
\begin{center}
\includegraphics[scale=0.15]{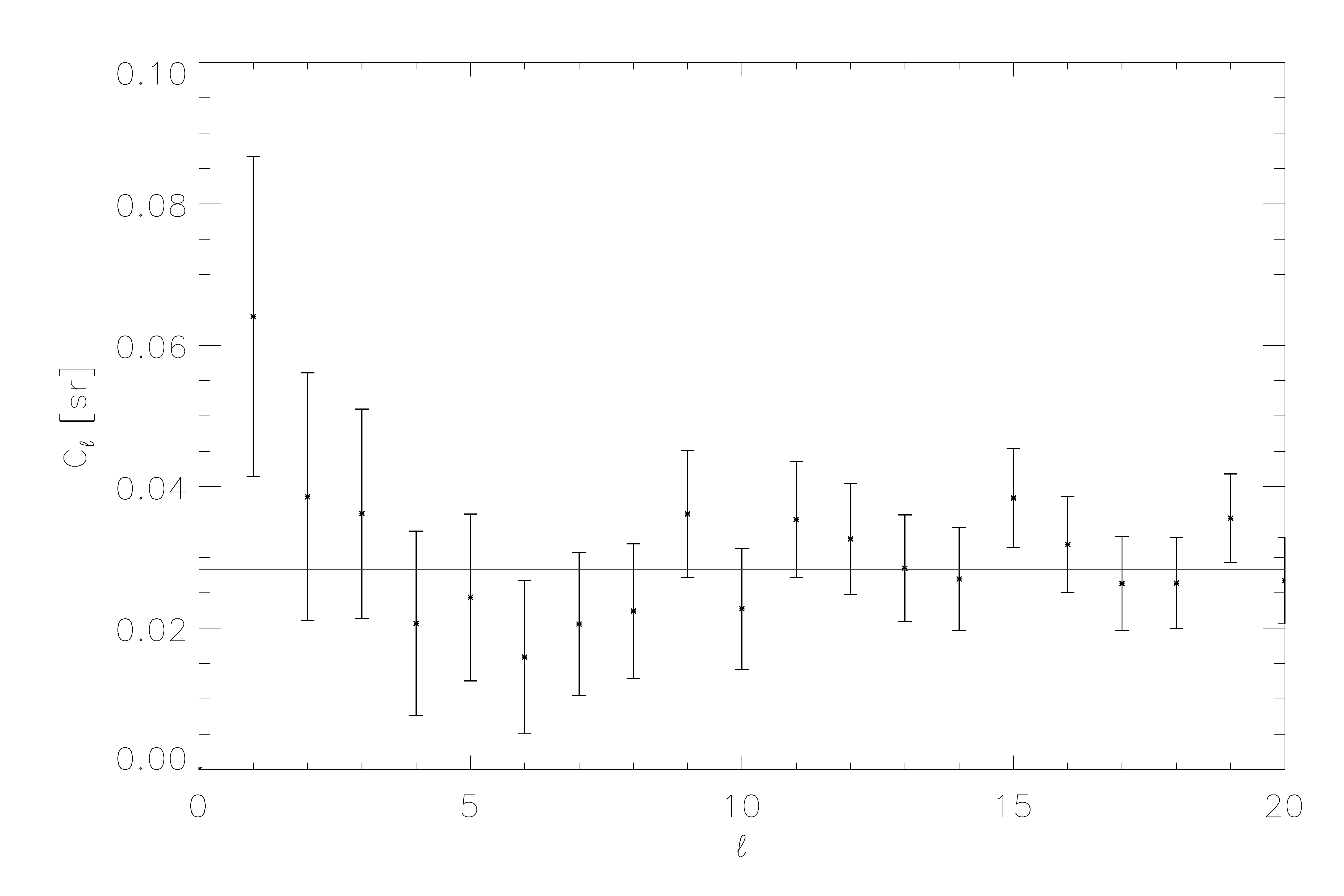}
\caption{X-SZ cross-correlation amplitude, $A_{\rm XSZ}$, power spectrum. The black sample presents the measured power spectrum from MILCA and RASS full-sky maps, the red line shows the prediction when isotropy is satisfied.}
\label{axszps}
\end{center}
\end{figure}

For high $\ell$ ($\ell > 1000$), we tested the isotropy of the cosmological parameters derived from the 416 individual regions on the sky. The X-SZ power-spectrum amplitude, $A_{\rm XSZ}$, strongly depends on cosmological parameters, $A_{\rm XSZ} \propto \sigma^8_8 \Omega^{3.5}_m$, but the shape of the power spectrum is only marginally affected by cosmology.
Thus, we used the best-fitting cosmology found in \citet{hur15} to predict the X-SZ cross power spectrum. For $1000 < \ell < 2000$, we computed $A_{\rm XSZ}$ in each region of the sky through a linear fit of the measured X-SZ cross power spectra (presented in Fig.~\ref{xszsp}) with the \citet{hur15} X-SZ power spectrum prediction.
Consequently, we derived a full-sky map of $A_{\rm XSZ}$ in Healpix $N_{\rm side} = 8$ pixellization. From this map, we computed the $A_{\rm XSZ}$ angular power spectrum and corrected it for partial sky coverage effects. \\

In Fig.~\ref{axszps} we present the power spectrum of $A_{\rm XSZ}$ up to $\ell = 20$. 
When isotropy is satisfied, all the fluctuations we should observe in the $A_{\rm XSZ}$ map should be due to noise and cosmic variance, and should be uncorrelated between independent regions of the sky. Consequently, in the isotropic situation we expect to observe white noise.
The dipole of $A_{\rm XSZ}$ is consistent with isotropy at 1.5$\sigma$. We observe that all modes up to $\ell = 20$ are consistent with isotropy for the large-scale structure matter distribution.\\

The $A_{\rm XSZ}$ power spectrum can be interpreted as a measurement of the isotropy of the Universe. Considering that the CMB shows that the matter distribution is isotropic at $z \simeq 1100$, the most natural solution for interpreting an eventual isotropic behavior of the large-scale structure at low-$z$ is through an anisotropic growth factor induced by large-scale dark energy anisotropies. Consequently, we translate our constraints on the $A_{\rm XSZ}$ dipole into constraints on the dark energy large-scale anisotropies.

\begin{figure}[!th]
\begin{center}
\includegraphics[scale=0.3,angle=90]{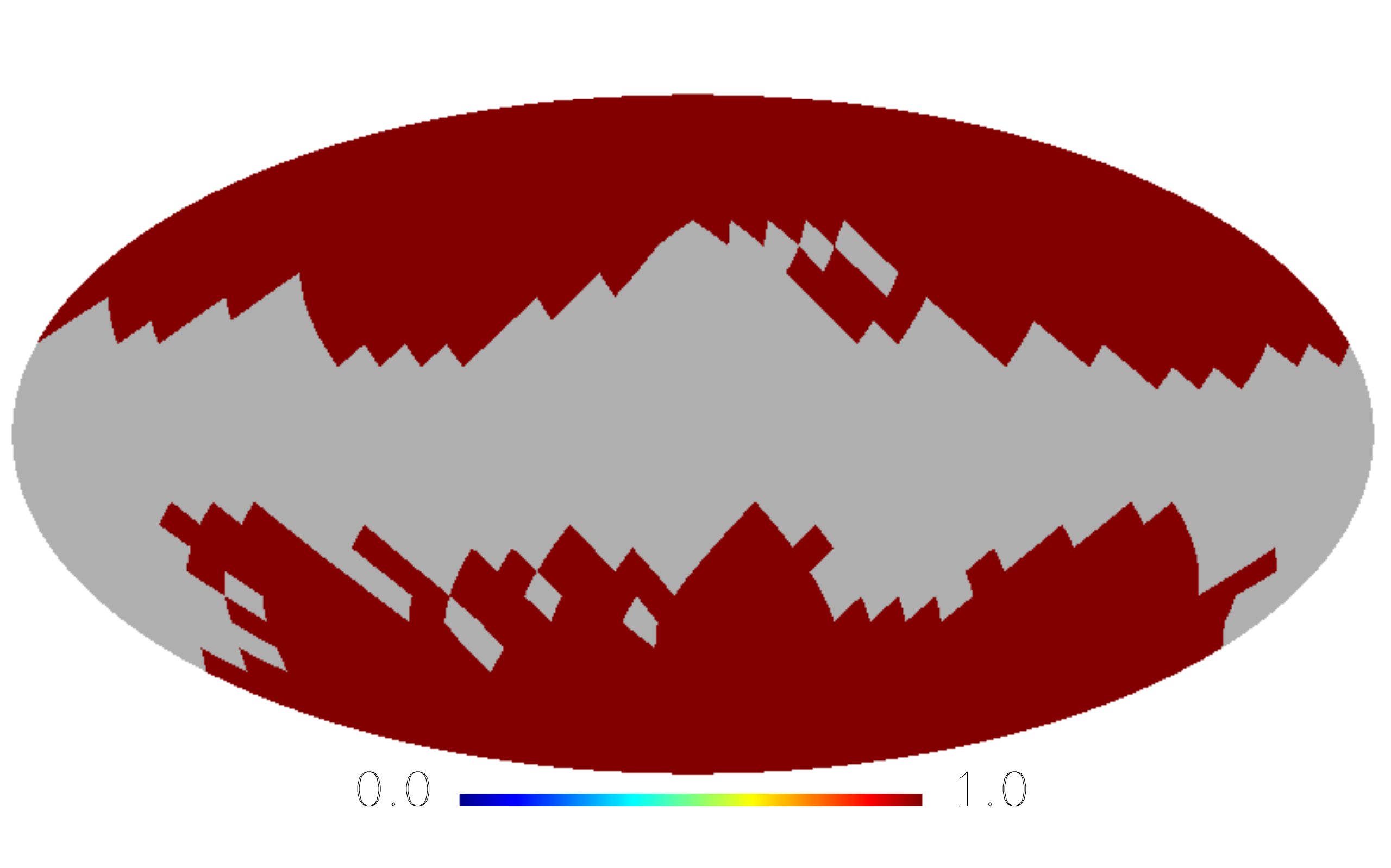}
\includegraphics[scale=0.3,angle=90]{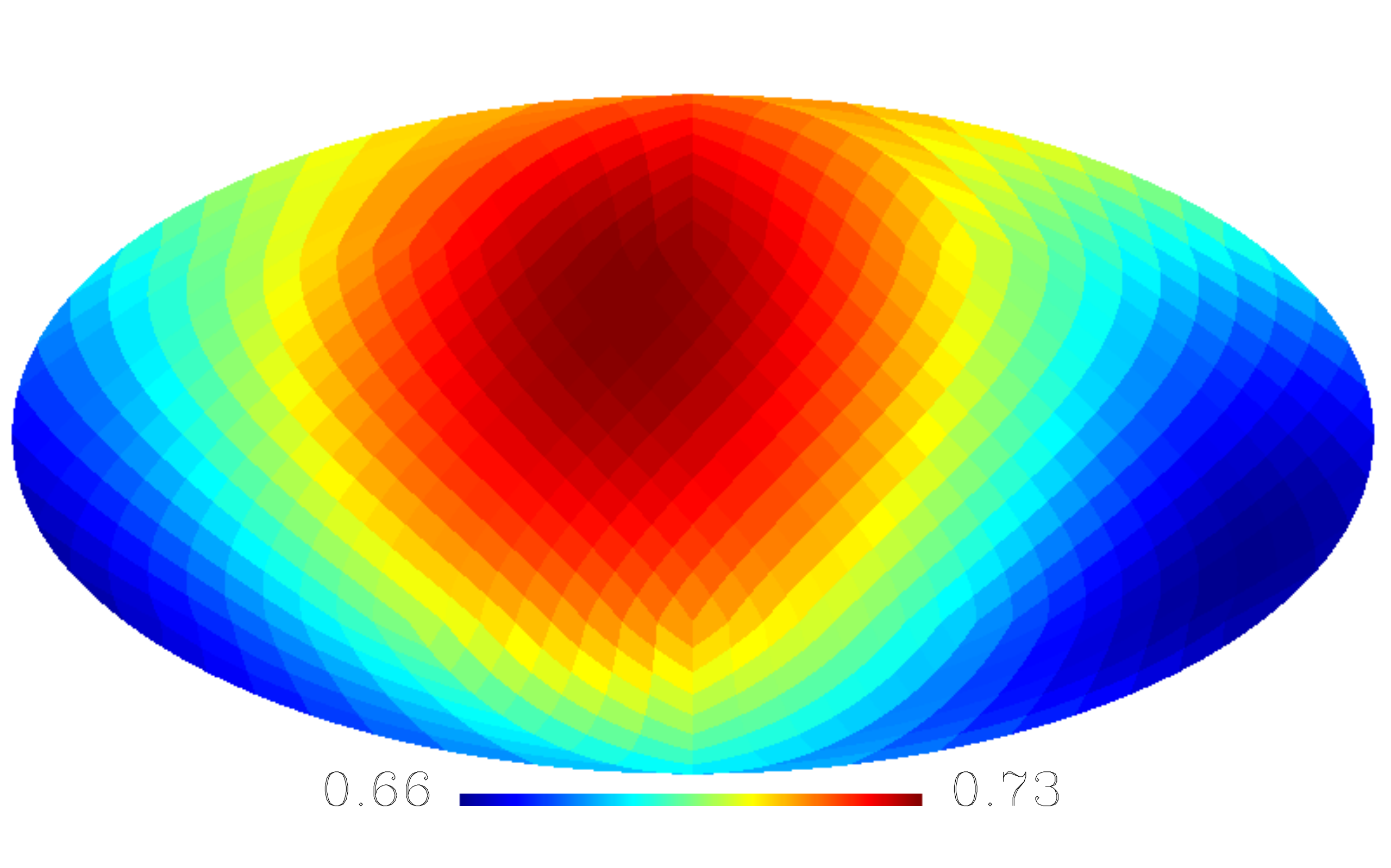}
\caption{Top panel: footprint of the sky regions used (in red) and rejected (in gray) from the analysis. Bottom panel: best-fitting dark energy dipole in Galactic coordinates.}
\label{dedip}
\end{center}
\end{figure}

In Fig.~\ref{dedip} we present the distribution on the sky of the 416 regions used in the analysis (top panel) and the best-fitting $\Omega_\Lambda$ dipole (bottom panel). To facilitate visualization, we display the full-sky best-fitting dipole. The maximum of this dipole is found in the direction $(l,b) = (22.5^o,24.6^o)$ with a significance of 1.5$\sigma$. Considering that the best-fitting dipole maximum is oriented near the Galactic center, we cannot exclude some residual foreground bias in the tSZ MILCA map. This dipole is also oriented toward regions masked by our cuts on the Galactic thermal dust emission, thus its orientation has to be considered carefully for potential residual bias and its very low level of significance.
Interpreted as an upper limit, we have $\Delta \Omega_\Lambda < 0.07$ at 95\% confidence level, with $\Delta \Omega_\Lambda$ being the dipole amplitude.

\section{Discussion and conclusion}

We have demonstrated that the measurement of a large-scale structure power-spectrum can be achieved on a small sky fraction ($f_{\rm sky} < 0.13$\%) that is limited by cosmic variance up to $\ell = 1000$.\\
We showed that the cosmic-variance-dominated part of the SZ-X power spectrum covariance matrix can be used to set cosmological constraints with the same accuracy level than the SZ-X or SZ  power spectra. However, owing to the similar degeneracies between cosmological parameters and astrophysical processes for the X-SZ power spectrum and the X-SZ-X-SZ tri-spectrum, such cosmological constraints do not allow us to break degeneracies between cosmology and astrophysical processes.
The derived constraints confirm previous results derived from galaxy number counts \citep[e.g.,][]{planckSZC} and power spectra \citep[e.g.,][]{planck15s,hur15}.\\
We also present a tight constraint on the dark energy dipole in the local Universe ($z < 1$), implying that $\Omega_\Lambda$ does not vary across the sky by more than 10\% at 95\% confidence level.
These constraints are competitive compared to those derived from the study of the distance modulus variation through SNIa \citep{ant10,lin16}.

\bibliographystyle{aa}
\bibliography{isotropy.bib}

\end{document}